\documentclass[twocolumn,showpacs,preprintnumbers,amsmath,amssymb]{revtex4}

\usepackage{graphicx}
\usepackage{dcolumn}
\usepackage{bm}

\begin{document}


\title{Asymmetric Fermi superfluid with different atomic species in a harmonic trap}

\author{C.-H. Pao}
\author{Shin-Tza Wu}
\affiliation{%
Department of Physics, National Chung Cheng University, Chiayi
621, Taiwan
}%

\author{S.-K. Yip}
\affiliation{ Institute of Physics, Academia
Sinica, Nankang, Taipei 115, Taiwan}%

\date{August 20, 2007}

\begin{abstract}
We study the dilute fermion gas with pairing between two species and
unequal concentrations in a harmonic trap using the mean field theory and the local
density approximation.
We found that the system can exhibit a superfluid shell structure sandwiched by the normal fermions.
This superfluid shell structure occurs if the mass ratio is larger then certain critical
value which increases from the weak-coupling BCS region to the strong-coupling BEC side.
In the strong coupling BEC regime, the radii of superfluid phase are less sensitive to the mass ratios and are similar to the case of pairing with equal masses.
However, the lighter leftover fermions are easier to mix with the superfluid core
than the heavier ones. A partially polarized superfluid can be found if the majority fermions are lighter, whereas phase separation is still found if they are heavier.
\end{abstract}

\maketitle

\section{\label{introduction}Introduction}

Recently experimental progress has raised strong interest in
studying the superfluid pairing in the ultracold Fermi gases
\cite{expr}. Through the Feshbach resonance \cite{Feshbach}, the
effective interaction between atoms can be tuned over a wide range.
This technique led us to investigate the crossover between the
condensation of weak-coupling Cooper pairs and the Bose-Einstein
condensation (BEC) of strong-coupling tightly bound pairs\cite{EL}.
More recently, two experimental groups extend a step further by
controlling the polarization of the two states Fermi gases
\cite{ketterle06} and open a new era for studying the imbalanced
Fermi gases, which is a topic related to many interesting areas from
condensed-matter physics, nuclear physics, to quark
matter \cite{Forbes05}.

Stimulated by these excellent experiments of the imbalanced Fermi
gases, there are intense theoretical works in the past two years. For a homogeneous system, the smooth crossover is known to be destroyed when the populations of
the fermions are unequal \cite{pao06,sheehy06,sheehy07,son06}.
In an inhomogeneous case, $e.g.$ in a
harmonic trap potential, phase separation occurs near resonance with
the superfluid is at the trap center surrounding by the normal
phase \cite{paoyip06,yi06,desilva06,haque06,chevy06,Lobo06}.
Finite temperature phase diagrams and density profiles are also studied in this
imbalanced system recently \cite{stoof06,sesarma07,levin06,simons07,chengyip07}.

Feshbach resonances between different atomic species have also been
reported \cite{FBumass} and the pairing of unequal masses fermions has
received theoretical interest. Focusing on the pairing of
$^6$Li-$^{40}$K mixtures, the phase diagram was studied both in a
homogeneous system \cite{Iskin06} and in a trap potential
\cite{lin06}. Other works on a general unequal
mass ratios, including the analytic results at unitary limit \cite{wu06}
and the evolution of the tricritical point \cite{parish07}
have also been studied recently.

In this paper, we study the
density profiles for unequal mass fermions in harmonic traps at zero temperature.
Within the local density approximation, or Thomas-Fermi approximation,
we solve the BCS gap equation self-consistently at fixed total number
of particles and polarization. We analyze the density profiles from unpolarized to highly imbalanced phases. Due to the large number of possible atoms (e.g.
$^2$H, $^3{\rm He}^\ast$ \cite{mcnamara06},
$^6$Li, $^{40}$K, $^{87}$Sr \cite{grimm}, and $^{173}$Yb \cite{fukuhara07}),
we study general mass ratios from 0.1 to 10 and
particular $(k_F a)^{-1} = -0.5$, $0$, and $1.0$ as
the representative scenarios for weak-couping BCS,
on resonance, and strongly paired BEC regimes.
We found, on the weak coupling BCS side,
the superfluid can form a shell structure sandwiched
between normal fermions for large mass ratios.
At resonance, the normal fluids can be either partially or
completely polarized, depending on the mass ratios.
On the BEC side, the radii of superfluid core are insensitive to the mass ratios.
A partially polarized superfluid can be found if the majority fermions
are lighter, whereas phase separation is still found if they are heavier.
We also report the axial density profiles of the population difference and discuss
the significant differences compared to the pairing with equal masses.

Mismatch Fermi surfaces may result a different ground state than phase separations,
in particular the so-called Fulde-Ferrel-Larkin-Ovchinnikov (FFLO) phase \cite{FFLO}. In this paper, we leave out this possible FFLO phase for future investigation \cite{yoshyip07}.

The remainder of this paper is organized as follows: In Sec. II
we briefly review the mean-field approximation for the dilute two states of fermion
atoms with unequal masses. We present our numerical results along with discussion,
in Sec. III. Finally, in Sec. IV, we conclude with a briefly summary.

\section{\label{form}Formalism:}

We start from the two-component fermion system across a wide Feshbach resonance
which may be described by an effective one-channel Hamiltonian:
\begin{eqnarray}
H& =& \label{eqh} \sum_{{\bf k}, \sigma} \xi_\sigma ({\bf k})
c^\dagger_{{\bf k},\sigma} c_{{\bf k},\sigma}\nonumber \\
& & +\ g \sum_{\bf
k,k^\prime, q} c^\dagger_{{\bf k+q},h} c^\dagger_{{\bf
k^\prime -q},l} c_{{\bf k^\prime},l} c_{{\bf k},h}\ ,
\end{eqnarray}
where $\xi_{\sigma}({\bf k})\, =\, \hbar^2 k^2/2m_\sigma - \mu_\sigma$, $g$ the bare
coupling strength, and
the index $\sigma$ runs over the two species ($h$ and $l$).
Within the BCS mean field approximation, the excitation
spectrum in a homogeneous system for each species is (see e.g.
\cite{WY03} for details)
\begin{eqnarray}
E_\sigma ({\bf k})& =& \frac{ \xi_{\sigma}({\bf k}) - \xi_{-
\sigma}({\bf k})}{ 2}\ +\nonumber \\ & & \sqrt{ \left ( \frac{ \xi_{\sigma}({\bf
k}) + \xi_{-\sigma} ({\bf k})} {2 } \right )^2\, + \, \Delta^2}\ .
\label{eqdisp}
\end{eqnarray}
where
$-h \equiv l$. In order to include the pairing between unequal masses,
we use $m_h$ ($m_l$) to represent the mass of each component with heavier
(lighter) mass and rewrite the dispersion relation of Eq. (\ref{eqdisp}) as
\begin{eqnarray}
E_{h,l}({\bf k})& =& \mp \left [ { \hbar^2 k^2 \over 4 m_r} { \gamma
-1 \over \gamma + 1}\, +\, h \right ]\ +\nonumber \\ & & \sqrt{ \left ({\hbar^2 k^2
\over 4 m_r}\, -\, \mu \right )^2 \, +\, \Delta^2 }\ ,
\label{eqdisp1}
\end{eqnarray}
with the reduced mass $m_r$, the mass ratio $\gamma = m_h/m_l$, the chemical potential
difference $h\, \equiv (\mu_h - \mu_l)/2$ and the average chemical potential
$\mu\, \equiv\, (\mu_h + \mu_l)/2$.

For the fermion gases in a harmonic trap, the system can be treated as
homogeneous locally if the number of particles are sufficiently
large.
Within this local density approximation, the densities
at position $\vec{ r}$ depends only on the local effective chemical
potentials
\begin{equation}
\mu_{h,l} (\vec{r}) \ =\ \mu_{h,l}^{0} - V_{h,l} (\vec{ r})
\end{equation}
where $ V_{h,l} (\vec{ r})$ is the trap potential for the
species $h$ or $l$.   To simplify our presentation
we shall consider an isotropic trap where the potential
depends only on the radius $r$.

The density profiles can be found via
\begin{eqnarray}
N_s(r)& = & N_h(r) + N_l (r)\, =\, \int { d^3 k \over
(2 \pi)^3} \nonumber \\
& & \left [ 1 + 2 \left ( { \hbar^2 k^2 \over 4 m_r}
\, -\,  \mu(r) \right ) { f(E_h) - f(-E_l) \over E_h +
E_l} \right  ]\ , \label{eqns}\\
N_d (r) & = & N_h (r) - N_l (r)\nonumber \\
& =& \int { d^3 k
\over (2 \pi)^3}\biggl [\,  f(E_h)\ -\ f(E_l) \,\biggr ]\ , \label{eqnd}
\end{eqnarray}
with $f$ is the Fermi function. The total number of particles
$N\, = \, \int d^3 r N_s(r) $
and the polarization of the system is
defined as
\begin{equation}
P \ \equiv\ {1 \over N}\int d^3 r N_d(r)\ .
\end{equation}
Note that the polarization is positive (negative) for a system with the majority are
heavier (lighter).

Now the pairing field $\Delta$ depends on position also. In the
local density approximation, it obeys an equation similar to the
homogeneous case \cite{pao06,WY03}:
\begin{eqnarray}
 - { m_r \over 2 \pi a} \Delta (r)& =& \Delta(r)
 \int \frac{ d^3 k}{ (2\pi)^3} \nonumber \\
& &\left [ \frac{ 1 - f(E_h) - f(E_l) }{ E_h +
E_l }\, -\, { 2m_r \over \hbar^2 k^2} \right ]\ .
\label{eqgap}
\end{eqnarray}
For a given s-wave scattering length $a$, we solve equations
(\ref{eqns}), (\ref{eqnd}) and (\ref{eqgap}) self-consistently for
fixed total number of particles $N$ and polarization $P$. The
solutions of the ``gap equation'' (\ref{eqgap}) may not be unique and
the physical solution is determined by the condition of minimum free
energy among the multiple solutions. The detailed procedure can be
found in reference \cite{paoyip06}.

To avoid extra complications, we confine ourselves
to the case where the trapping potential is harmonic.
We further assume that it is
identical for the two species.  Thus
$V (r) = \frac{1}{2} \alpha_{h,l} r^2$ with
$\alpha_h = \alpha_l = \alpha$.   This would occur,
for example, if both species are trapped magnetically
and if the fermions have identical magnetic moments,
or when the fermions are trapped optically with
different lasers of appropriate intensities and detunings.
In this case, within the local density approximation,
the density profiles for the same polarization
$P$ but different total particle numbers are related to
each other via simple scaling.
This allow us to present our results in a manner
which is independent of the total particle numbers (see below).
When the trap potentials are unequal,  many other
scenarios can occur.  We shall leave the investigation
of this more complicated case to the future.

\section{\label{result}results}

In this section, we investigate the density profiles for various
polarizations, mass ratios and coupling strengths from positive detuning BCS
superfluid to negative detuning BEC side. With the aid of density
profiles, we plot the superfluid phase diagram in a harmonic trap for different
mass ratios. Finally, we also evaluate the axial density profiles
of the population difference and discuss the effect due to the mass ratios
at different coupling strengths.

\begin{figure}
\includegraphics[width=\columnwidth]{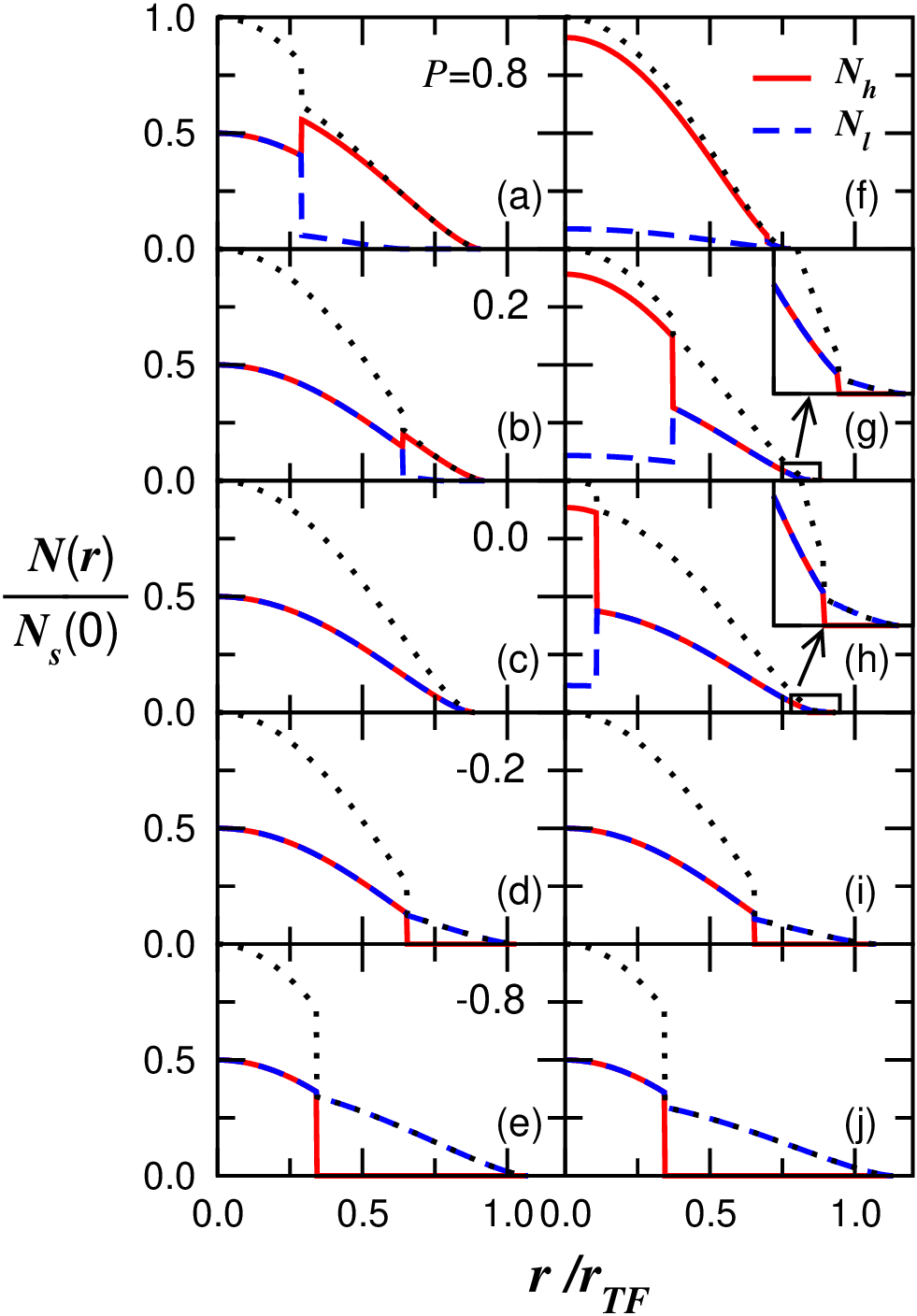}
\caption{(Color online)
The radial density
profiles for heavy fermions (solid lines) and light fermions (dashed
lines) at the unitary limit ($(k_F a)^{-1} = 0$). The dotted lines
are the total density profiles. The mass ratios $\gamma$ are equal
2.0 for plots (a)-(e) and 5.0 for plots (f)-(j). The vertical scale is
normalized to the total density of fermions at trap center
($N_s(0)$). $r_{TF}$ is the Thomas-Fermi radius of the normal
majority cloud.} \label{fig1g0}
\end{figure}

\begin{figure}
\includegraphics[width=\columnwidth]{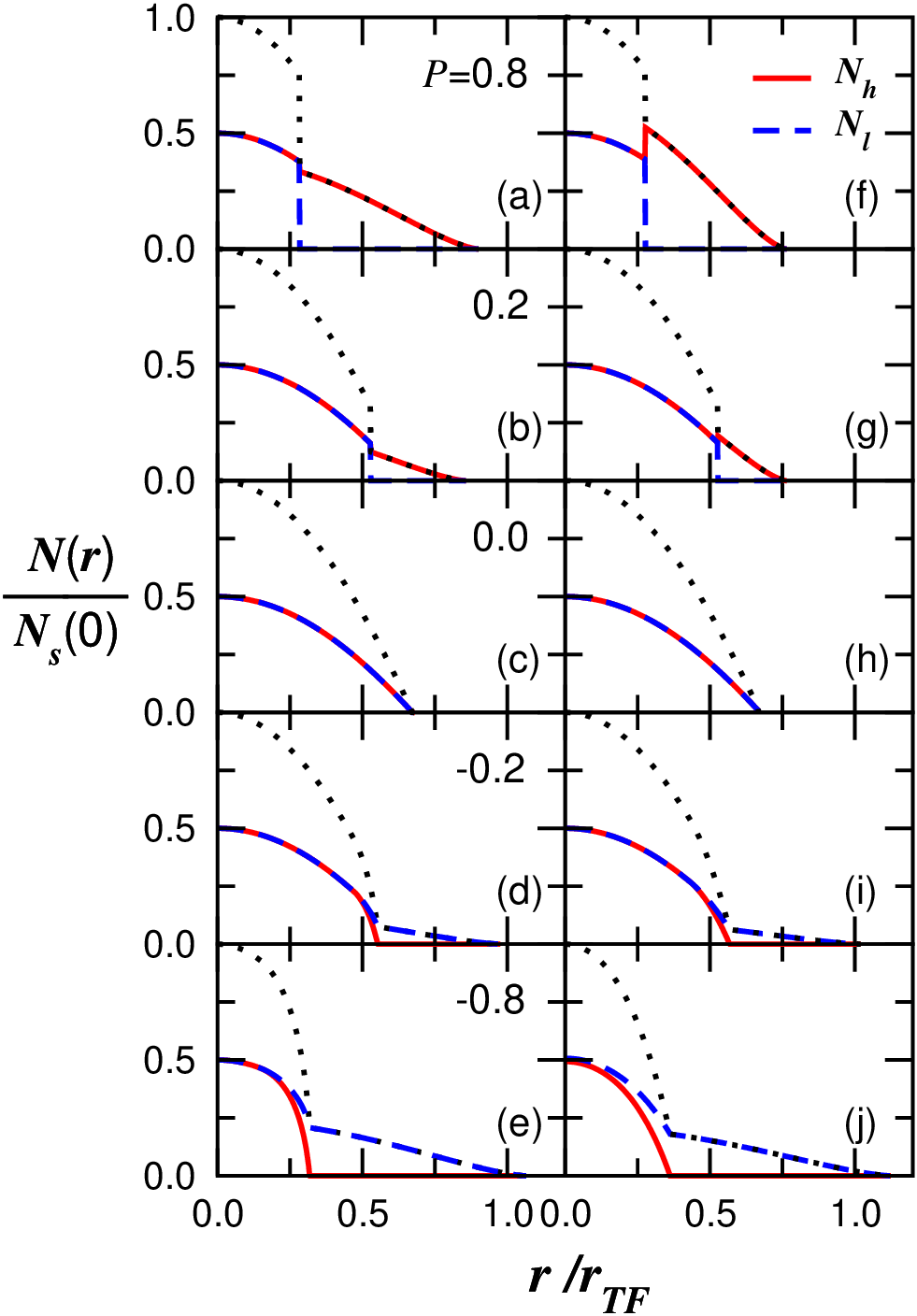}
\caption{(Color online)
The radial density
profiles on the BEC side with $(k_F a)^{-1} = 1.0$. The parameters
are the same as in Fig. 1} \label{fig2g1}
\end{figure}

\begin{figure}
\includegraphics[width=\columnwidth]{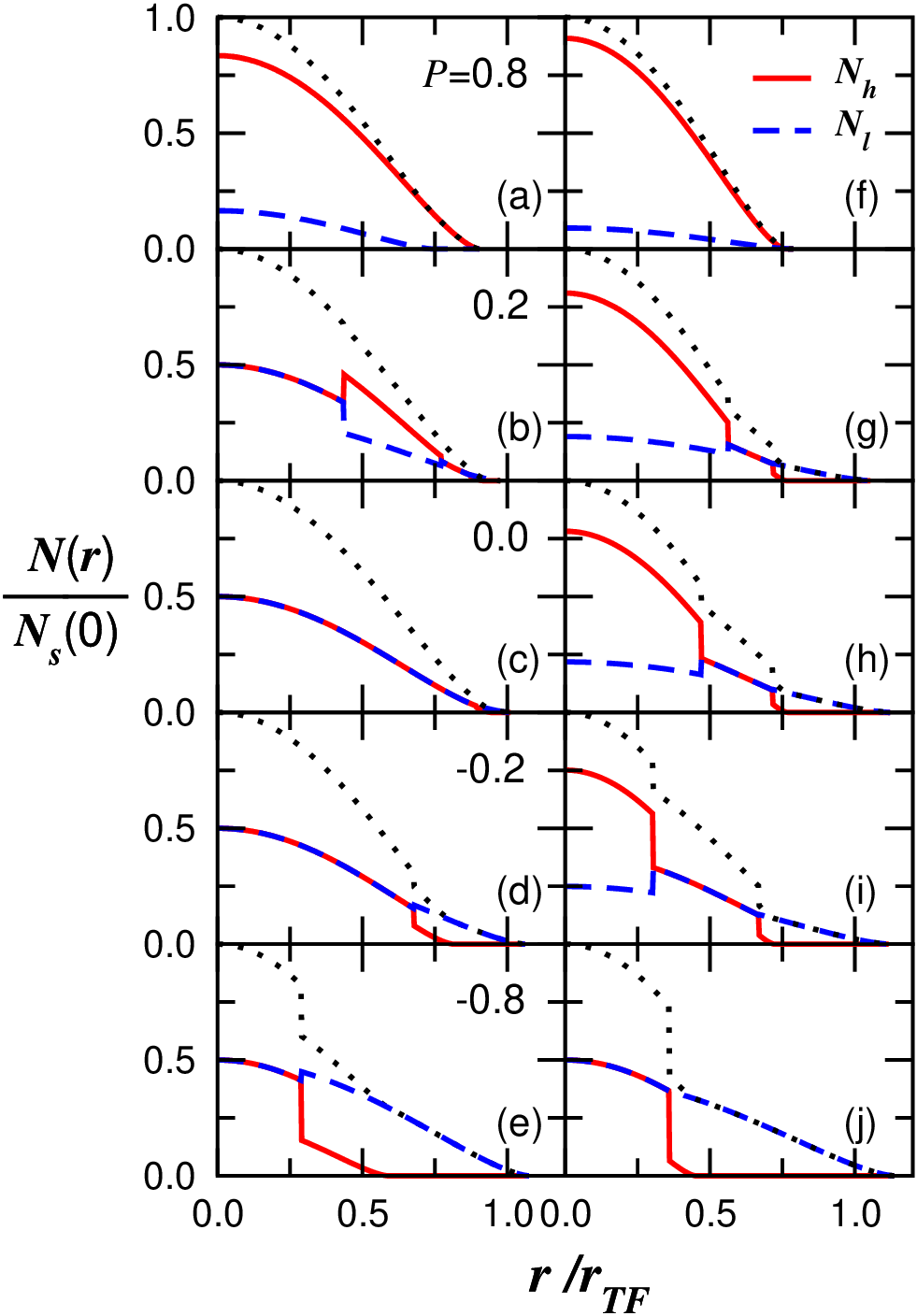}
\caption{(Color online)
The radial density
profiles on the BCS side with $(k_F a)^{-1} = -0.5$. The parameters
are the same as in Fig. 1} \label{fig3g-0.5}
\end{figure}

We begin by examining the radial density profiles for the mass
ratios $\gamma\, =\, 2.0$, and $5.0$ with various polarizations from
0.8 (majority are heavier) to -0.8 (majority are lighter).
The densities $N(r)$ are normalized to the total density at the trap center and distances are normalized to the Thomas-Fermi radius of the non-interacting majority gas of the same atom number.
In Figs. \ref{fig1g0}--\ref{fig3g-0.5}, we plot the radial density
profiles at three different coupling strengths with $(k_F a)^{-1} = $ 0
(the unitary limit), 1.0 (strong coupling BEC), and -0.5
(weak-coupling BCS).

In the unitary limit, the system exhibits a superfluid core
surrounded by normal fermions for $\gamma = 2.0$ [Fig.
\ref{fig1g0}(a)--(e)]. The density profiles are qualitatively the
same between positive and negative polarizations.
For example, for the large polarizations
$P = \pm 0.8$ [Fig. \ref{fig1g0}(a) and (e)], we found that the
superfluid core of the majority are lighter is only slightly larger
than that of the majority are heavier.
However, for
$P > 0$ there are small amounts of minority in the normal region but it is
completely polarized for the case of $P < 0 $.
The above results are in accordance with Fig. 2 of reference [\onlinecite{wu06}].
At the interface between the superfluid and the normal
fluid, the two phases are in equilibrium and the local
polarization of the latter phase should be given by the same
value obtained from our earlier bulk calculations there.
The asymmetry between the positive and negative polarizations becomes more obvious for
larger mass ratio, $e.g.$, $\gamma = 5$ in Fig.
\ref{fig1g0}(f)--(j). In Fig. \ref{fig1g0}(g) and (h), the
superfluid is sandwiched between two regions with normal fermions.
The inner core is a normal state with heavy particles being the majority,
whereas the outer normal shell consists purely the lighter species
\cite{note1}. The superfluid shell extends toward the trap center as
the polarization decreases from large positive value. At equal
populations [Fig. \ref{fig1g0}(h)], we found the system is still
phase separated into normal polarized fermions and unpolarized
superfluid regions. This is quite different compare to the equal or
small mass ratio ($e.g.$, $\gamma =2.0$ here) cases which the system
is superfluid for the entire trap \cite{note2}.

The existence of this normal fluid core can be understood
as due to the fact that this normal fluid has
a {\em higher} number density than the paired superfluid
at this mass ratio \cite{wu06}, and hence it
``sinks'' towards the center of the trap where the
trap potential is the minimum (recall that we have assumed $\alpha_h = \alpha_l$).
We note also here that, as shown in Fig \ref{fig1g0},
the total number density is a decreasing function of
$r$.
In fact, the results of \cite{wu06} then indicate that
the critical mass ratio above which for the occurrence
of a normal state core with a superfluid shell is
$\gamma^* = 3.9$ at resonance (see Fig. 3 of reference [\onlinecite{wu06}]).

In Fig. \ref{fig2g1}, we plot the
radial density profiles for $\gamma = 2.0$ and 5.0 with $(k_F
a)^{-1} = 1.0$. At this coupling strength, the radii of the
superfluid core are found to be only very weakly dependent
on the mass ratio or the sign of polarization.
However, the phases are very different
depending on whether the majority particles are the
heavier or lighter ones.
We found that all particles are paired for a unpolarized case ($P = 0$).
When 
the majority fermions are heavier ($P > 0$),
the system contains an unpolarized
superfluid core with a surrounding outside shell
of normal fluid consisting of the majority
particles alone.  Thus phase separation occurs in this case
between the completely paired superfluid and the completely polarized
normal phase.  On the other hand, when the majority is light
($P < 0$), the equally paired superfluid core at the trap center is
surrounded by a partially polarized superfluid mixture
phase, with a normal shell occupying the outermost region of the trap.
This normal shell is again found to be completely polarized.
In this strongly paired BEC regime, all minority particles
exist only within the bound pairs and thus within the superfluid
phase(s) only.   Note also that the coupling strength is
now strong enough so that the superfluid always has a
higher density than the normal phase, and the superfluid
shell structure found at resonance no longer appears here.

In the BCS regime, since the pairing interaction
is weak,  the density profiles are most easily understood
by first considering the normal state density profiles.
We note here that due to the different mass ratios and
unequal populations, these two profiles have very different
dependence on the position $r$. In the presence of
the weak interaction, pairing would occur only
at locations where the local populations are almost equal,
or when the density is sufficiently large so that
the effective interaction is sufficiently strong.
This results in a rich structures in the density profiles
as we shall show.
In Fig. \ref{fig3g-0.5}, we plot the radial density
profiles for the coupling strength $(k_F a)^{-1} = -0.5$ and mass
ratios $\gamma = 2.0$ and 5.0. The symmetry between the
positive and negative polarization is lost even for $\gamma = 2.0$
[Fig. \ref{fig3g-0.5}(a)--(e)]. At large populations
imbalances ($e.g.$, $P = \pm 0.8$ in Fig.
\ref{fig3g-0.5}), a superfluid core still forms for majority are
lighter but only normal fermions exist in the trap for majority are
heavier. Similar to the unitary limit, the system exhibits a
superfluid core surrounded by normal fermions for a wide range of
polarization at $\gamma = 2.0$. But at larger $\gamma$ [Fig.
\ref{fig3g-0.5}(f)--(j)], the heavier normal fermions prefer to stay
in the trap center and the superfluid phase is outside
this normal phase. We also found that the superfluid shell grows as
the population of lighter fermions increase [Fig.
\ref{fig3g-0.5}(g)--(i)]. This superfluid shell structure finally
reaches the trap center and all normal fermions are forced into the
outside shell. We would like to emphasize that the shell structure
of the normal fermions are not the same as the cases in the unitary
limit. In Fig. \ref{fig1g0}(g) and (i), the outside shell of the
normal fermions are fully polarized but both components of fermions
exist here for this weak-coupling side.

\begin{figure}
\includegraphics[width=\columnwidth]{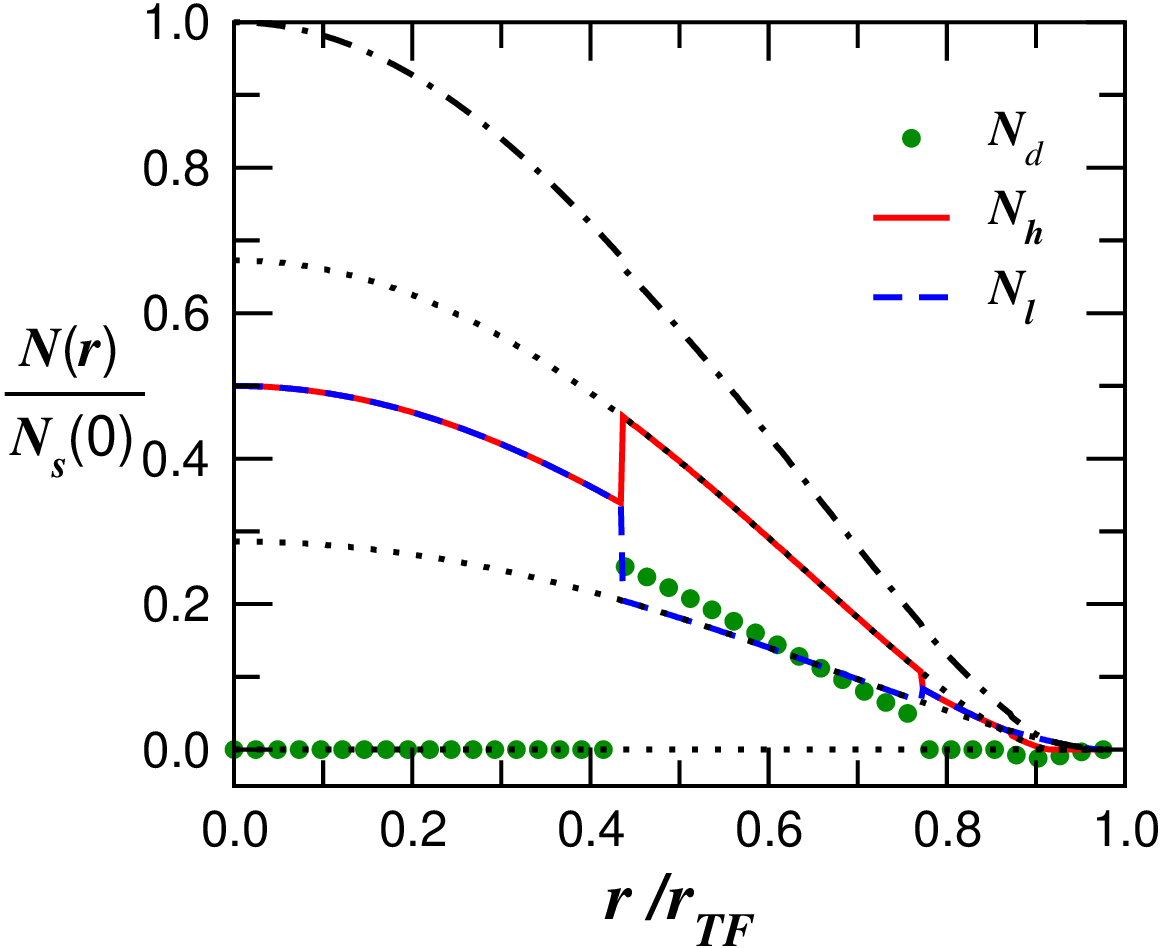}
\caption{(Color online)
The radial density profiles for heavy fermions (solid lines) and light fermions (dashed
lines) with parameters $(k_F a)^{-1} = -0.5$, $\gamma = 2.0$, and $P
= 0.2$. The dotted lines are extrapolated density profiles of each component
as if they were non-interacting.
Solid circles represent the difference
between the two components of fermions. Note that, $N_d$ is slightly
less than zero near the edge of the trap.} \label{fig4}
\end{figure}

In Fig. \ref{fig3g-0.5}(b), we found that the density profile
contains two regions of superfluid and polarized fermions. As mentioned,
a stable superfluid occurs when either the density of each
component of fermion is large or the size of Fermi surfaces of these two
fermion states are approximately equal.
In Fig. \ref{fig4}, we enlarge the
Fig. \ref{fig3g-0.5}(b) and plot the density profiles of both
components of the non-interacting fermions also. It shows that the
superfluid shell just occurs near the crossing of these two
non-interacting density profiles since the superfluid gap is
easier to open near this point \cite{lin06}. On the other hand, the density is
large enough such that the superfluid
phase is again stable at the trap center.

\begin{figure}
\includegraphics[width=\columnwidth]{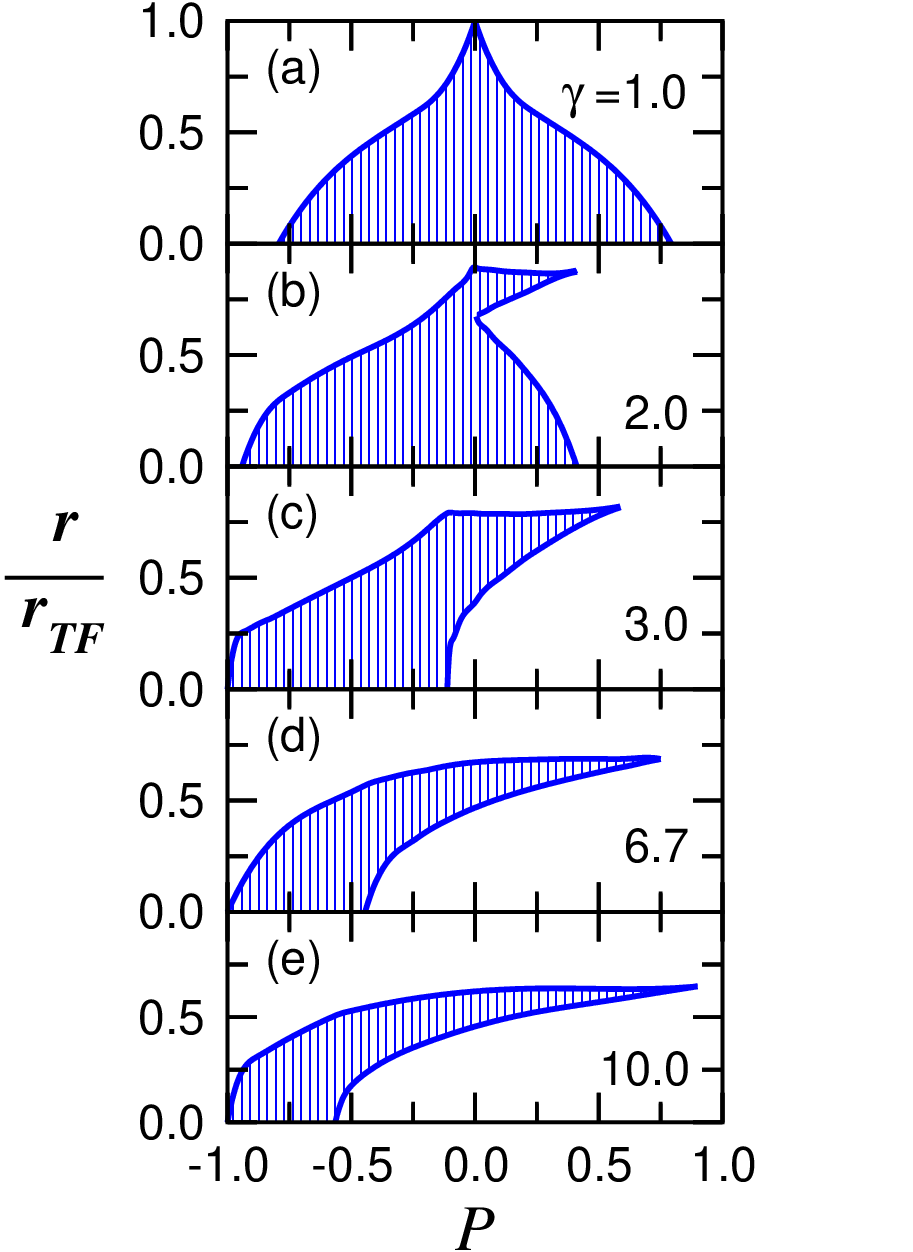}
\caption{(Color online)
Superfluid phase diagram for $(k_F a)^{-1} = -0.5$ in the trap. The shaded area represents the superfluid phase.}
\label{fig5}
\end{figure}

In view of the rich structure on the BCS side, we turn next to the
phase diagram of the superfluid in a trap potential in this regime.
We plot the phase diagrams in Fig. \ref{fig5} for several different
mass ratios with the same coupling strength $(k_F a)^{-1} = -0.5$.
In Fig. \ref{fig5}(a) ($\gamma = 1$), the superfluid phase extends
to the entire trap at the unpolarized system ($P = 0$) and is symmetric
about this point. For $\gamma > 1$, the fermion paired between
different masses, the maximum radius of the superfluid phase moves
toward negative polarization where the majority of the system are
lighter. The symmetric phase diagram does not hold anymore. At
increasing mass ratio ($\gamma \gtrsim 3$), we found that the
superfluid at the trap center is not stable for the positive
polarization. Instead, the superfluid phase occurs at some finite
radius for the parameters we studied in Fig.
\ref{fig5}(b)--(e). The shell structure of the superfluid is present
for a wide range of polarizations as the mass ratio $\gamma$
increases. In Fig. \ref{fig5}(b), $\gamma = 2.0$, the superfluid
phase is split into two regions by the normal fermions and vice
versa for $0 \lesssim P \lesssim 0.4$, corresponding to the
concentric structure we discuss in Fig. \ref{fig3g-0.5}(b) above.
Note that, the mass ratios in Fig. \ref{fig5}(b), (c), and (d)
correspond to the mixtures of $^3$He*--$^6$Li, $^2$H--$^6$Li and
$^6$Li--$^{40}$K.

\begin{figure}
\includegraphics[width=\columnwidth]{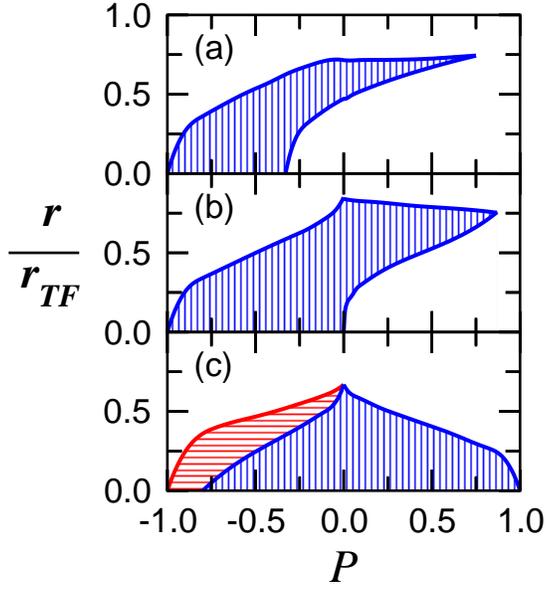}
\caption{(Color online)
Superfluid phase diagram for $(k_F a)^{-1} = -0.5$ (a), 0.0 (b) and 1.0 (c). The mass ratio $\gamma = 5.0$. The vertical shaded area shows unpolarized superfluid phase and the horizontal shaded area (in plot (c)) shows partially polarized superfluid phase. }
\label{fig6}
\end{figure}

In Fig. \ref{fig6}, we also plot the superfluid phase for three different
coupling strengths with $\gamma = 5.0$. The superfluid shell structure still
exists at the unitary limit [Fig. \ref{fig6}(b)] but only for positive
polarizations. On the BEC side [Fig. \ref{fig6}(c)], the symmetry of the
superfluid phase about the equal population almost regains. However,
there is a region where the superfluid and leftover fermions are
mixed with each other for $P < 0$. For $P > 0$,
there is only unpolarized superfluid phase and the system is phase separated.

\begin{figure}
\includegraphics[width=\columnwidth]{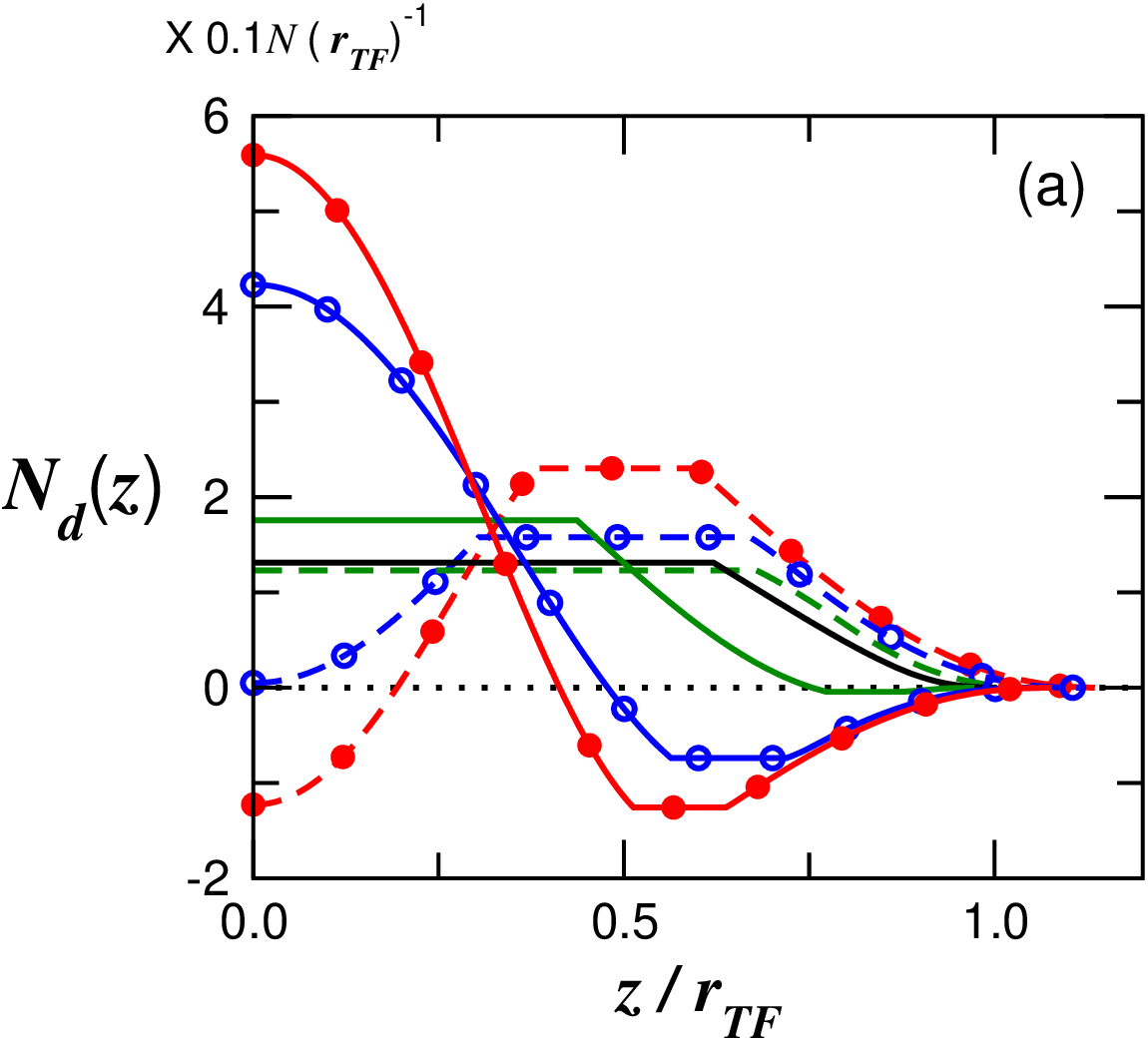}
\includegraphics[width=\columnwidth]{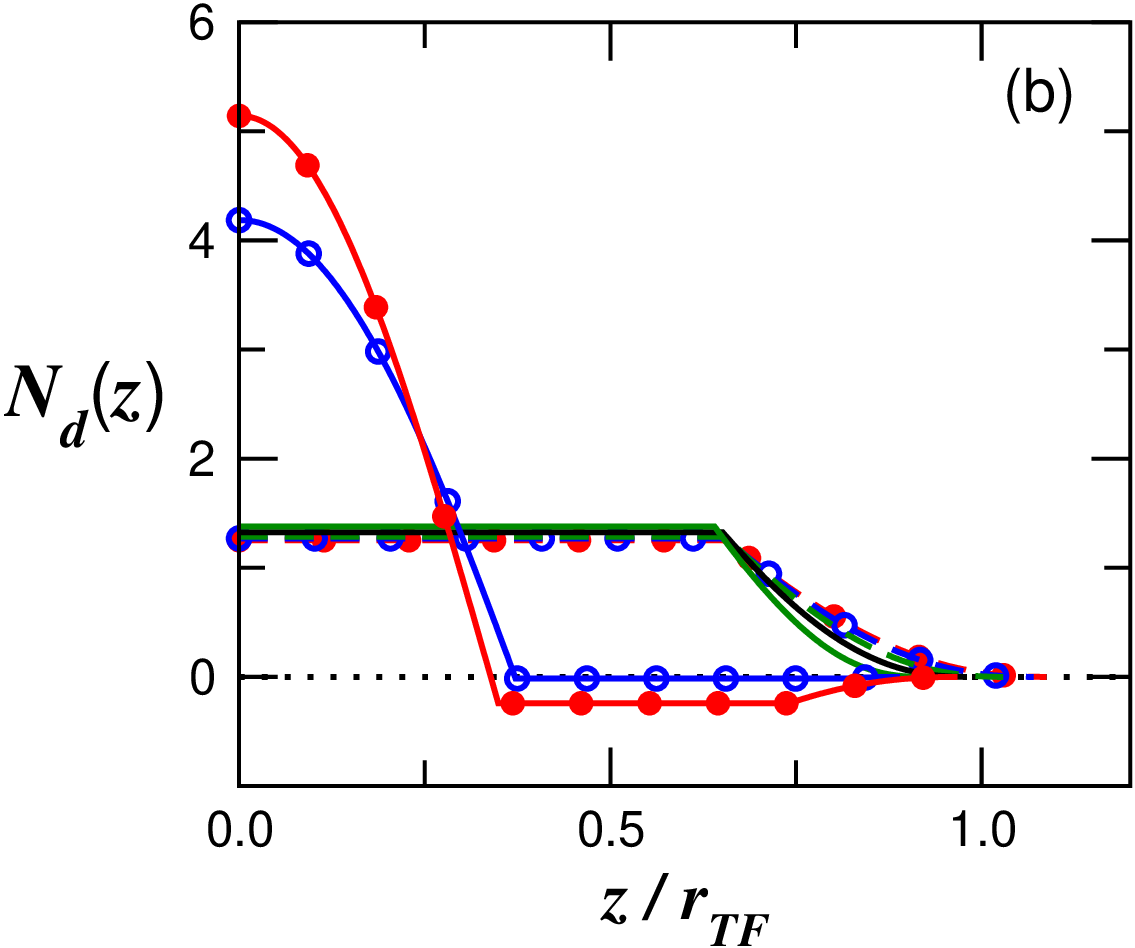}
\includegraphics[width=\columnwidth]{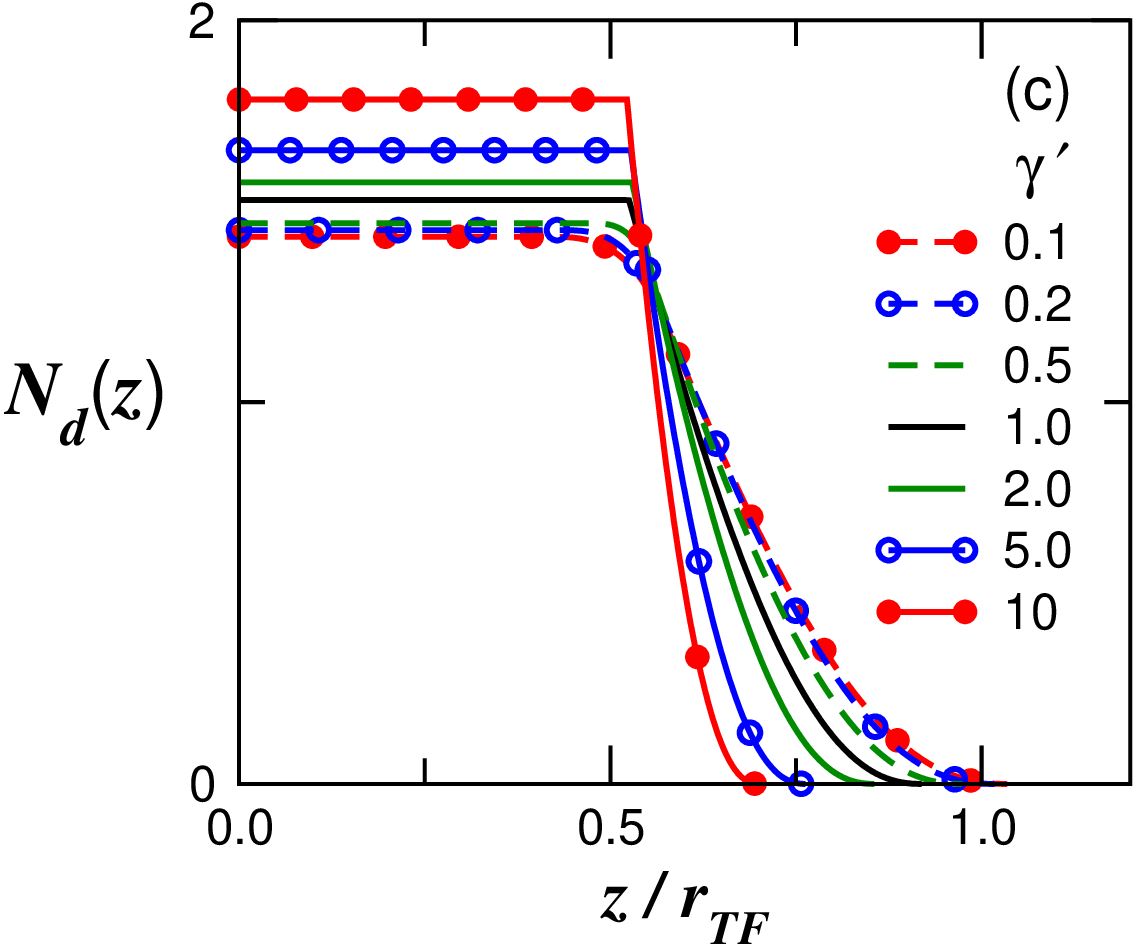}
\caption{(Color online)
Axial density profiles for $(k_F a)^{-1} = -0.5$ (a), 0.0 (b) and 1.0 (c) with $|P| = 0.2$. Solid lines correspond to the majority heavier ($\gamma^\prime > 1$) and dashed lines corresponding to the majority lighter ($\gamma^\prime < 1$).}
\label{fig7}
\end{figure}

Lastly we would like to examine the axial density profiles of the
population difference defined as \cite{paoyip06}:
\begin{equation}
N_d(z) \ =\ \int dx dy \bigl [ N_h(\vec{r}) - N_l (\vec{r}) \bigr ]\ .
\end{equation}
In Fig. \ref{fig7}, we plot these results for
$(k_F a)^{-1} = -0.5,\, 0.0$ and $1.0$ and the mass ratios
($\gamma^\prime \equiv m_{majoity} /m_{minority}$) between majority and minority species from $0.1$ to $10.0$ at polarization $|P| = 0.2$.

On the BCS side [Fig. \ref{fig7}(a)] and the unitary limit [Fig. \ref{fig7}(b)],
the horizontal segments of each curve $N_d (z)$ correspond to
the unpolarized superfluid. For the cases of majority lighter
($\gamma^\prime \lesssim 0.2$), we found that the minority (with heavier
mass) crowd into the trap center and have higher population than the
majority locally. Thus, the axial density profiles show the dip structure
in the trap center. In the other limit, $\gamma^\prime \gtrsim 5.0$ (the majority are heavier), the peak of the axial density profile appears right at the trap center.

At the unitary limit [Fig. \ref{fig7}(b)], the
axial density profiles are similar except for $\gamma^\prime = 5.0$ and
$10.0$. This is the results where the heavier fermions have
larger populations near the trap center and the superfluid shows a shell structure as we discuss in Fig. \ref{fig1g0}.

On the strong coupling BEC side [Fig. \ref{fig7}(c)], the structures of all
of the curves are similar for different mass ratios. The radii of
the superfluid core in this coupling strength are not sensitive to
$\gamma^\prime$ either. The only difference is that $N_d(z)$ are smooth
at the phase boundary for the cases of the majority are lighter
but there exist kinks for the cases of the majority are
heavier . It is consistent with the results in Fig.
\ref{fig2g1} that the lighter fermions are easier to mix with
superfluid than the heavier ones.

\section{\label{con}Conclusion}

We have studied the two-component fermion system with
unequal masses and populations across Feshbach resonance.
For the pairing with unequal species,
the superfluid can be sandwiched by the normal fermions and form a superfluid shell structure. This superfluid shell structure is easier to observe on the weak-coupling BCS side or at the unitary limit with large mass ratios ($i.e.$, in the systems of $^{40}$K--$^{173}$Yb, $^6$Li--$^{40}$K, or $^6$Li--$^{87}$Sr\cite{grimm}). For a given mass ratios, this superfluid shell extends toward the trap center as the population of lighter fermions increase.

In the strong coupling BEC regime, the superfluid phase is less sensitive to the
mass ratios and is similar to the case of pairing with equal masses.
However, lighter leftover fermions are easier to penetrate into the superfluid core
than the heavier ones. At coupling strength $(k_F a)^{-1} = 1.0$, phase separation between the completely polarized superfluid and the completely polarized normal phase occurs for the majority are heavier. On the other hand, the equal paired superfluid core at the trap center is surrounded by a partially polarized superfluid mixture phase and a normal shell occupying the outmost region of the trap.

\begin{acknowledgments}
This research was supported by the National Science Council of
Taiwan under grant numbers NSC95-2112-M-194-011 (CHP), NSC95-2112-M-194-003 (STW) and
NSC95-2112-M-001-054 (SKY).
\end{acknowledgments}


\begin{thebibliography}{99}

\bibitem{expr}
C. A. Regal, M. Greiner, and D.S. Jin, Phys. Rev. Lett. {\bf 92},
040403 (2004); M. W. Zwierlein {\em et al.}, {\it ibid.} {\bf 92},
120403 (2004); J. Kinast, S.L. Hemmer, M.E.Gehm, A. Turlapov, and J.E. Thomas, {\it ibid.} {\bf 92}, 150402 (2004); T. Bourdel {\em et al}, {\it ibid.} {\bf 93},
050401 (2004); C. Chin {\em et al.}, Science {\bf 305}, 1128
(2004); and references therein.

\bibitem{Feshbach} E. Tiesinga, B. J. Verhaar, H. T. C. Stoof,
Phys. Rev. A {\bf 47}, 4114 (1993); S. Inouye {\em et al.}, Nature
{\bf 392}, 151 (1998); P. Courteille, R.S. Freeland, D.J. Heinzen, F.A. van Abeelen, and
B.J. Verhaar, Phys. Rev. Lett. {\bf 81}, 69 (1998); J. L. Roberts {\em et al.}, {\em ibid.}
{\bf 81}, 5109 (1998).

\bibitem{EL}
D. M. Eagles, Phys. Rev. {\bf 186}, 456 (1969); A. J. Leggett in
{\em Modern Trends in the theory of condensed matter}, edited by
A. Pekalski and J. Przystawa (Springer-Verlag, Berlin, 1980).

\bibitem{ketterle06} M.W. Zwierlein, A. Schirotzek, C.H. Schunck, and
W. Ketterle, Science {\bf 311}, 492 (2006); G. B. Partridge, W. Li,
R.I. Kamar, Y. Liao and R.G. Hulet {\it ibid.} {\bf 311}, 503
(2006); Y. Shin, M.W. Zwierlein, C.H. Schunck, A. Schirotzek, and W.
Ketterle, Phys, Rev. Lett. {\bf 97}, 030401 (2006); C.H. Schunck, Y.Shin, A. Schirotzek, M.W. Zwierlein, and and W. Ketterle, Science {\bf 316}, 867 (2007); G.B. Partridge,
Wenhui Li, Tean-an Liao, R.G. Hulet, M. Haque, and H. T. C. Stoof,
{\sl ibid.} {\bf 97} 190407 (2006).

\bibitem{Forbes05} MMichael McNeil Forbes,
E. Gubankova, W. V. Liu, and Frank Wilczek,
Phys. Rev. Lett. {\bf 94}, 017001 (2005); R. Casalbuoni
and G. Nardulli, Rev. Mod. Phys. {\bf 76}, 263 (2004) and references therein.

\bibitem{pao06}
C.-H. Pao, Shin-Tza Wu, and S.-K. Yip, Phys. Rev. B{\bf 73}, 132506
(2006); {\bf 74}, 189901(E) (2006).

\bibitem{sheehy06}
D.E. Sheehy and L. Radzihovsky, Phys. Rev. Lett. {\bf 96}, 060401
(2006).

\bibitem{sheehy07}
D.E. Sheehy and L. Radzihovsky, Ann. Phys. in press
(cond-mat/0607803).

\bibitem{son06}
D. T. Son and M. A. Stephanov, Phys. Rev. A{\bf 74}, 013614 (2006).

\bibitem{paoyip06}
C.-H. Pao and S.-K. Yip, J. Phys. : Condens. Matter {\bf 18}, 5567
(2006).

\bibitem{yi06} W. Yi and L.M. Duan Phys. Rev. A {\bf 73}, 031604(R) (2006).

\bibitem{desilva06} T.N. De Silva and E.J. Mueller, Phys. Rev. A{\bf 73}, 051602(R)
(2006); Phys. Rev. Lett. {\bf 97}, 070402 (2006).

\bibitem{haque06} M. Haque and H.T.C. Stoof, Phys. Rev. A{\bf 74}, 011602(R) (2006); Phys. Rev. Lett. {\bf 98}, 260406 (2007).

\bibitem{chevy06} F. Chevy, Phys. Rev. Lett. {\bf 96}, 130401 (2006).

\bibitem{Lobo06} C. Lobo, A. Recati, S. Giorgini and S. Stringari, Phys. Rev. Lett. {\bf 97}, 200403 (2006).

\bibitem{stoof06}
K.B. Gubbels, M.W.J. Romans, and H.T.C. Stoof, Phys. Rev. Lett. {\bf
97}, 210402 (2006).

\bibitem{sesarma07} R. Sensarma, W. Schnedier, R. B. Diener and M. Randeria,
  e-print arXiv:cond-mat/0706.1741.

\bibitem{levin06}
Chih-Chun Chien, Qijin Chen, Yan He, and K. Levin, Phys. Rev. Lett.
{\bf 97}, 090402 (2006); {\sl ibid.} {\bf 98}, 110404 (2007); Qijin
Chen, C.A. Regal, D.S. Jin, and K. Levin, Phys. Rev. A {\bf 74},
011601(R) (2006).

\bibitem{simons07} M.M. Parish, F.M. Marchetti, A. Lamacraft, and B.D. Simons,
Nature Physics 3, 124 (2007).

\bibitem{chengyip07}
Chi-Ho Cheng and S. K. Yip, Phys. Rev. B {\bf 75}, 014526 (2007).


\bibitem{FBumass}
C. A. Stan, M. W. Zwierlein, C. H. Schunck, S. M. F. Raupach, and W.
Ketterle, Phys. Rev. Lett. {\bf 93}, 143001 (2004); S. Inouye, J.
Goldwin, M. L. Olsen, C. Ticknor, J. L. Bohn, and D. S. Jin, {\sl
ibid.} {\bf 93}, 183201 (2004).

\bibitem{Iskin06} M. Iskin and C. A. R. Sa de Melo,
Phys. Rev. lett. {\bf 97}, 100404 (2006).

\bibitem{lin06} G.-D. Lin, W. Yi, and L.-M. Duan, Phys. Rev. A{\bf 74}, 031604(R)
(2006).

\bibitem{wu06}
Shin-Tza Wu, C.-H. Pao, and S.-K. Yip, Phys. Rev. B{\bf 74} 224504
(2006).

\bibitem{parish07}
M.M. Parish, F.M. Marchetti, A. Lamacraft, and B.D. Simons, Phys.
Rev. Lett. {\bf 98}. 160402 (2007).

\bibitem{mcnamara06}J. M. McNamara, T. Jeltes, A. S. Tychkov, W. Hogervorst and W. Vassen, Phys. Rev. Lett. {\bf 97}, 080404 (2006).

\bibitem{grimm} Current project at University of Innsbruck: Heteronuclear molecules of fermionic lithium, potassium and strontium (FeLiKx). Please see website at $<$http://www.uibk.ac.at/exphys/ultracold/$>$

\bibitem{fukuhara07}T. Fukuhara, Y. Takasu, M. Kumakura and Y. Takahashi,
 Phys. Rev. Lett. {\bf 98}, 030401 (2007).

\bibitem{FFLO}
P. Fulde and R. A. Ferrell, Phys. Rev. {\bf 135}, A550 (1964); A. I.
Larkin and Yu. N. Ovchinnikov, Zh. \'{E}ksp. Teor. Fiz. {\bf 47},
1136 (1964) [Sov. Phys. JETP {\bf 20}, 762 (1965)].

\bibitem{yoshyip07}
Studies of the FFLO phase in the imbalanced Fermi gas can be found in recent papers $e.g.$, N.Yoshida and S.-K. Yip, Phys. Rev. A{\bf 75}, 063601 (2007) and 
T. Mizushima, M. Ichioka, and K. Machida, cond-mat/0705.3361, and references therein.

\bibitem{WY03}
S.-T. Wu and S.-K. Yip, Phys. Rev. A {\bf 67}, 053603 (2003).

\bibitem{note1}
This sandwiched behavior was also observed for the case of
$^6$Li-$^{40}$K mixtures, e.g. in reference \cite{lin06}.

\bibitem{note2}
Fig 4. of reference \cite{wu06} missed the structure discussed here.

\end{thebibliography}

\newpage



\end{document}